\documentclass[conference,compsoc]{IEEEtran}
\IEEEoverridecommandlockouts
\usepackage{float}
\usepackage{amsmath,amsfonts}
\usepackage{algorithmic}
\usepackage{graphicx}
\usepackage{textcomp}
\usepackage{xcolor}
\usepackage{multirow}
\usepackage{microtype}
\usepackage{hyperref}
\usepackage{xurl}
\usepackage{bm}

\newcommand{\maybe}[1]{\textcolor{cyan}{}}
\newcommand{\code}[1]{\texttt{#1}}

\newcommand{\toremove}[1]{\textcolor{blue}{}}

\newcommand{\mycomment}[1]{}
\newcommand{\artifactLink}[1]{https://anonymous.4open.science/r/AGNOMIN-4D44/}

\def\BibTeX{{\rm B\kern-.05em{\sc i\kern-.025em b}\kern-.08em
    T\kern-.1667em\lower.7ex\hbox{E}\kern-.125emX}}
\begin{document}
\pagestyle{plain} 
\title{AGNOMIN - \underline{A}rchitecture A\underline{gno}stic \underline{M}ulti-Label Funct\underline{i}on \underline{N}ame Prediction
}
\author{
    \IEEEauthorblockN{Yonatan~Gizachew~Achamyeleh$^1$, Tongtao~Zhang$^2$,  Joshua~Hyunki~Kim$^1$, Gabriel~Garcia$^1$, Shih-Yuan~Yu$^1$,\\ Anton~Kocheturov$^2$,     
    Mohammad~Abdullah~Al~Faruque$^1$}
    {\textit{$^1$Dept. of Electrical Engineering and Computer Science, University of California, Irvine, CA, USA}}
    \\
    {\textit{\{yachamye, joshuhk6, gegarci1, shihyuay, alfaruqu\}@uci.edu}} 
    \\
    {\textit{$^2$Siemens Technology, Princeton, NJ, USA, \{tongtao.zhang, anton.kocheturov,\}@siemens.com}}
}

\maketitle

\begin{abstract}
Function name prediction is crucial for understanding stripped binaries in software reverse engineering, a key step for \textbf{enabling subsequent vulnerability analysis and patching}.
However, existing approaches often struggle with architecture-specific limitations, data scarcity, and diverse naming conventions.
We present AGNOMIN, a novel architecture-agnostic approach for multi-label function name prediction in stripped binaries.
AGNOMIN builds Feature-Enriched Hierarchical Graphs (FEHGs), combining Control Flow Graphs, Function Call Graphs, and dynamically learned \texttt{PCode} features.
A hierarchical graph neural network processes this enriched structure to generate consistent function representations across architectures, vital for \textbf{scalable security assessments}.
For function name prediction, AGNOMIN employs a Ren\'ee-inspired decoder, enhanced with an attention-based head layer and algorithmic improvements.

We evaluate AGNOMIN on a comprehensive dataset of 9,000 ELF executable binaries across three architectures, demonstrating its superior performance compared to state-of-the-art approaches, with improvements of up to 27.17\% in precision and 55.86\% in recall across the testing dataset.
Moreover, AGNOMIN generalizes well to unseen architectures, achieving 5.89\% higher recall than the closest baseline. 
AGNOMIN's practical utility has been validated through security hackathons, where it successfully aided reverse engineers in analyzing and patching vulnerable binaries across different architectures.

\end{abstract}

\begin{IEEEkeywords}
Function name prediction, multi-label classification, graph neural networks, and software reverse engineering.
\end{IEEEkeywords}

\section{Introduction}
\label{sec:intro}

Ensuring software integrity is paramount, especially for critical infrastructures often reliant on legacy systems. These systems, frequently lacking continuous updates or original development support, pose significant security risks~\cite{powner2016information, alexandrova2015legacy}. Initiatives like DARPA's AMP, HARDEN, and V-SPELL projects underscore the urgent need for advanced solutions to secure these vital digital assets~\cite{darpa_amp, darpa_harden, darpa_vspell}, motivating the drive towards automated vulnerability remediation.

Software Reverse Engineering (SRE) is crucial for analyzing and fortifying such systems~\cite{nsfghidra, hex2017ida, radare2book}. However, the common practice of stripping symbol tables from binaries creates a major obstacle: the loss of function names and semantic information~\cite{keliris2019icsref}. 
This is particularly critical in security contexts, where understanding function purposes is essential for automated vulnerability remediation, security auditing, and binary patching. 
This loss severely impacts reverse engineers who heavily rely on such textual information during initial analysis phases~\cite{votipka2020observational,mantovani2022re}, especially when identifying and updating deprecated cryptographic implementations or applying targeted security patches in stripped binaries.

Recent research has sought to address this issue by leveraging machine learning (ML) models for function name prediction~\cite{lacomis2019dire, he2018debin,david2020neural, Gao2021,SymLM,patrickevans2022xfl, patrickpuntrip,asmdepictor, yu2023cfg2vec}. 
However, these approaches are hindered by several key limitations: \textit{architecture dependence}, \textit{data scarcity}, and \textit{limited handling of diverse naming conventions}. 
Many existing models rely on architecture-specific features, which limits their applicability across diverse CPU architectures. 
For example,~\cite{lacomis2019dire, he2018debin,david2020neural, Gao2021,SymLM,patrickpuntrip, asmdepictor} demonstrate significant performance degradation when models trained on one architecture are applied to binaries from another. 
This architecture dependence is particularly problematic for scalable automated vulnerability assessment, where malicious code and vulnerable software often appear across multiple architectures.
Furthermore, the performance of these models often suffers due to data scarcity, especially for less common architectures or unique function names with insufficient training data. As highlighted in~\cite{maie_retargeted_architecture}, limited data availability leads to decreased model accuracy for unseen binaries, where models struggle to generalize beyond the training data distribution.  
Additionally, existing models exhibit difficulty in predicting names with high variability, including unique or underrepresented function names commonly encountered in diverse software environments.

To bridge this critical semantic gap and empower automated security analyses, we present AGNOMIN, a novel architecture-agnostic approach for multi-label function name prediction in stripped binaries.
AGNOMIN leverages Feature-Enriched Hierarchical Graphs (FEHGs), combining Control Flow Graphs, Function Call Graphs, and dynamically learned \texttt{PCode} features to represent binary functions comprehensively.
\texttt{PCode} is an intermediate representation used by the Ghidra SRE tool~\cite{nsfghidra} to represent the disassembled binary code in a more human-readable and architecture-independent format~\cite{ghidra_pcode}.
By utilizing \texttt{PCode} features, AGNOMIN can capture semantic information that is consistent across different CPU architectures.
This architecture-agnostic design is pivotal not only for name prediction but also for directly aiding vulnerability remediation by, for instance, identifying matching functions between stripped vulnerable binaries and patched reference binaries, a capability demonstrated in our case study (Sec.~\ref{sec:case_study}).

A hierarchical graph neural network (GNN) processes the enriched data to produce consistent functional representations across CPU architectures.
For function name prediction, AGNOMIN employs a decoder inspired by Ren\'ee~\cite{renee2023}, enhanced with an attention-based head layer and algorithmic improvements to address the specific challenges of multi-label function name prediction in an architecture-agnostic setting.
The attention-based head layer enables the decoder to dynamically focus on the most relevant aspects of the function embeddings for each label prediction, capturing the complex relationships and dependencies between labels. 
By integrating this enhanced decoder into our end-to-end framework, we enable accurate and informative predictions across diverse architectures and naming conventions.

We evaluate AGNOMIN on a comprehensive dataset of 396,096 functions in 9,000 ELF binaries across three architectures.
Our experimental results show that AGNOMIN significantly outperforms current state-of-the-art (SOTA) approaches in function name prediction across diverse architectures.
Compared to \texttt{XFL}~\cite{patrickevans2022xfl}, AGNOMIN achieves up to 27.17\% higher precision. Against \texttt{SymLM}~\cite{SymLM}, AGNOMIN demonstrates up to 11.88\% higher precision and 55.86\% higher recall.
These improvements can be attributed to AGNOMIN's architecture-agnostic approach, which leverages graph-based representations and learned \texttt{PCode} features to capture more relevant semantic information and generate accurate function name predictions.

Moreover, AGNOMIN's architecture-agnostic design allows it to generalize effectively to unseen architectures. When trained on \texttt{amd64} and \texttt{armel} binaries and evaluated on the previously unseen \texttt{i386} architecture, AGNOMIN achieves {3.12\%} higher precision and 5.89\% higher recall than the closest competitor, \texttt{CFG2VEC}~\cite{yu2023cfg2vec}. 
Its practical utility has been demonstrated in security hackathons, where reverse engineers used AGNOMIN to analyze and patch vulnerable binaries across architectures.
In one case, it helped identify deprecated cryptographic implementations in heavily optimized binaries, demonstrating its effectiveness in real-world security applications.
The main contributions are:

(1) AGNOMIN, an architecture-agnostic approach for multi-label function name prediction in stripped binaries.

(2) Evaluation of AGNOMIN on a dataset of 396,096 functions in 9,000 ELF binaries across three architectures, demonstrating its superior performance compared to SOTA methods~\cite{patrickevans2022xfl, SymLM}.

(3) An analysis highlighting AGNOMIN's ability to generalize to unseen architectures, showcasing {5.89\%} higher recall than \texttt{CFG2VEC}~\cite{yu2023cfg2vec} when tested but not trained with the \texttt{i386} binaries.

(4) Insights into the effectiveness of the attention-based head layer in the decoder, which enables AGNOMIN to capture complex relationships between labels and generate accurate predictions of function names.

(5) Evaluation of AGNOMIN in real-world security applications through DARPA-sponsored hackathons, demonstrating its effectiveness in vulnerability remediation tasks.


\section{Background and Related Works}
\label{sec:background}

\subsection{Software Reverse Engineering  Challenges}
Software Reverse Engineering (SRE) is essential for securing legacy systems by enabling the analysis and enhancement of outdated programs to prevent security breaches and system failures~\cite{vdurfina2013psybot, brumley2013native}.
However, SRE faces a significant challenge when function names and other semantic information are lost during the decompilation phase due to the removal of symbol tables from the binary~\cite{he2018debin}. 
This loss of meaningful identifiers hinders reverse engineers who rely heavily on such information during the initial analysis stages~\cite{votipka2020observational,mantovani2022re}, emphasizing the need for automated techniques to infer original function names accurately.

\subsection{Function Name Prediction Approaches}

Recent research tackles function name prediction in stripped binaries using ML-based methods, which fall into three main categories:

\subsubsection{Multi-class Classification} Early approaches treated function name prediction as a multi-class classification problem, mapping each function to a name from a predefined list. For example, Debin predicts function names based on a probability distribution across known names~\cite{he2018debin}. 
\texttt{CFG2VEC} offered a distinct perspective by employing function embeddings to deduce names based on structural similarity~\cite{yu2023cfg2vec}. 

However, these approaches faced limitations in scalability and adaptability due to their reliance on a static set of names, making them ineffective for binaries with unseen function names or diverse naming conventions.
For example, \texttt{CFG2VEC} must process and store embeddings for every function in the comparison space, posing significant challenges regarding memory efficiency, cost and practical scalability.
Additionally, the fixed namespace approach exacerbated the challenges posed by data sparsity, as unique naming conventions across different codebases result in many individual function names occurring rarely or only once within training datasets. 
The sparsity issue complicates the learning process, as models fail to learn how to predict sparse labels.

\subsubsection{Sequence-to-Sequence (seq2seq) Models}
These models treat the prediction task as transforming input sequences into output sequences of tokens, allowing them to generate new function names not seen during training. Examples include \texttt{NERO}~\cite{david2020neural}, \texttt{NFRE}~\cite{Gao2021}, \texttt{SymLM}~\cite{SymLM} \& \texttt{Asmdepictor}~\cite{asmdepictor}. While these models offer flexibility, they face challenges in scaling to large label spaces and predicting low-frequency tokens within function names~\cite{vaswani2017attention}.

\begin{figure*}[!ht]
    \centering
    \includegraphics[clip, width=.950\linewidth]{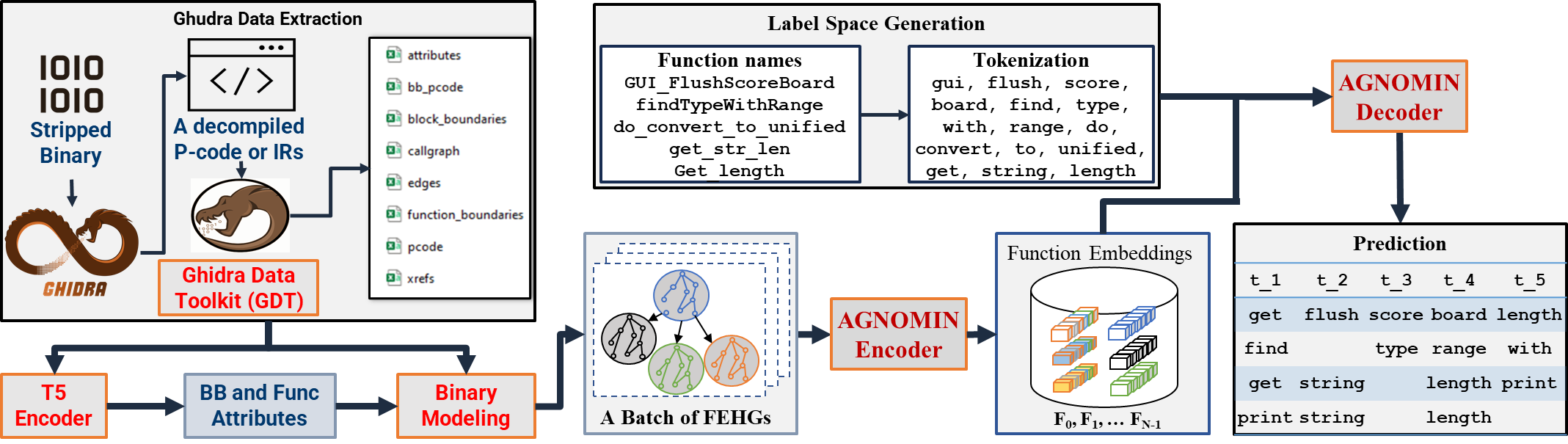}
    \caption{Overview of AGNOMIN’s design approach.}
    \label{fig:AGoG}
\end{figure*}

\subsubsection{Multi-Label Classification} The most recent approach frames function name prediction as a multi-label classification (MLC) problem. This approach decomposes function names into meaningful labels representing specific semantic aspects. This allows models to generate more descriptive and accurate names by capturing the complex relationships between different semantic aspects of a function~\cite{pfastrexml2016_jain}. \texttt{XFL}~\cite{patrickevans2022xfl} is a notable work that employs MLC for function name prediction, leveraging the \texttt{PfastreXML}~\cite{pfastrexml2016_jain} extreme multi-label classification library. 

\subsection{Architecture-Specific vs. Architecture-Agnostic}
\label{sec:related_work}

Existing approaches can also be categorized based on their ability to handle different CPU architectures:

\subsubsection{Architecture-Specific Approaches}
Most current approaches mainly focus on a single architecture, being trained and optimized for inference from that specific architecture.
Some examples include~\cite{lacomis2019dire, he2018debin,david2020neural, Gao2021,SymLM,patrickpuntrip, asmdepictor}. While these works can reconstruct source-level information effectively for their target architecture, their performance degrades significantly when applied to binaries from other architectures. This limitation restricts their usefulness in scenarios involving cross-compiled programs deployed across multiple architectures.

\subsubsection{Architecture-Agnostic Approaches}
Architecture-agnostic function name prediction is invaluable for debugging and malware detection. To our knowledge, \texttt{CFG2VEC}~\cite{yu2023cfg2vec} is the only approach supporting this.
However, \texttt{CFG2VEC} is limited in generalizing to unseen architectures. Moreover, \texttt{CFG2VEC} has a heavyweight inference procedure. To perform function name inference, it must compare a function against the embedding of every name in the set of ground truth function names, making it impractical for large-scale applications.
AGNOMIN addresses these challenges through its novel architecture-agnostic approach, combining Feature-Enriched Hierarchical Graphs, attention-based decoding, and algorithmic enhancements for multi-label classification.
\color{red}
\mycomment{
\subsection{Challenges in Function Name Prediction}
Despite the advancements in function name prediction, several key challenges remain:

\subsubsection{Architecture Dependence} Many existing models rely on architecture-specific features, limiting their applicability across diverse CPU architectures.
\subsubsection{Data Scarcity} The performance of these models often suffers due to limited data availability, especially for less common architectures or unique function names with insufficient training data.
\subsubsection{Diverse Naming Conventions} Existing models exhibit difficulty in predicting names with high variability, including unique or underrepresented function names commonly encountered in diverse software environments.
\subsubsection{Scalability} As the number of potential function names grows, many approaches struggle to scale efficiently, particularly when dealing with large codebases.
\subsubsection{Generalization} Achieving high precision in function name prediction often comes at the cost of recall, and vice versa. Striking a balance between these metrics while maintaining semantic accuracy remains a significant challenge.

AGNOMIN addresses these challenges through its novel architecture-agnostic approach, combining Feature-Enriched Hierarchical Graphs, attention-based decoding, and algorithmic enhancements for multi-label classification.
}

\color{black}

\section{Overview of the Design Approach}
\label{sec:overview}
Before diving into the technical details of AGNOMIN, we first present a high-level overview of our approach and its key components. AGNOMIN is developed as a plugin for Ghidra~\cite{nsfghidra}, aiming to enhance the efficiency of SRE processes. 
At its core, AGNOMIN utilizes a novel representation called Feature-Enriched Hierarchical Graphs (FEHGs), which combine Control Flow Graphs (CFGs), Function Call Graphs (FCGs), and dynamically learned \texttt{PCode} features. 

As shown in Fig.~\ref{fig:AGoG}, AGNOMIN's pipeline begins with data extraction and preparation via Ghidra, using the Ghidra Data Toolkit~\cite{yu2023cfg2vec} to efficiently extract essential information such as \texttt{PCode} and call graphs. 
The extracted \texttt{PCode} sequences are then processed using a pre-trained T5 model~\cite{DBLP:journals/corr/abs-1910-10683} to enhance the dataset with meaningful semantic attributes. 

A hierarchical GNN processes the enriched data to produce consistent functional representations across different CPU architectures, enabling AGNOMIN to overcome the limitations of architecture-specific models. These consistent representations are crucial for analyzing security vulnerabilities across different platforms and architectures.

For function name prediction, AGNOMIN employs a decoder inspired by Ren\'ee~\cite{renee2023}, incorporating attention-based head layers to focus on the most relevant aspects of the embeddings for label prediction.
Algorithmic enhancements, such as label pruning, further improve the decoder's handling of diverse label space.

The architecture-agnostic nature of AGNOMIN's embeddings enables additional security applications beyond name prediction, such as identifying similar functions between stripped and reference binaries across-architectures during vulnerability remediation tasks.

\section{Architecture-Agnostic Function Embedding Framework}
\label{sec:embedding}
\subsection{Modeling Binaries}
\begin{figure}[!t]
    \centering
    \includegraphics[trim=2pt 20pt 2pt 8pt,clip, width=.990\linewidth]{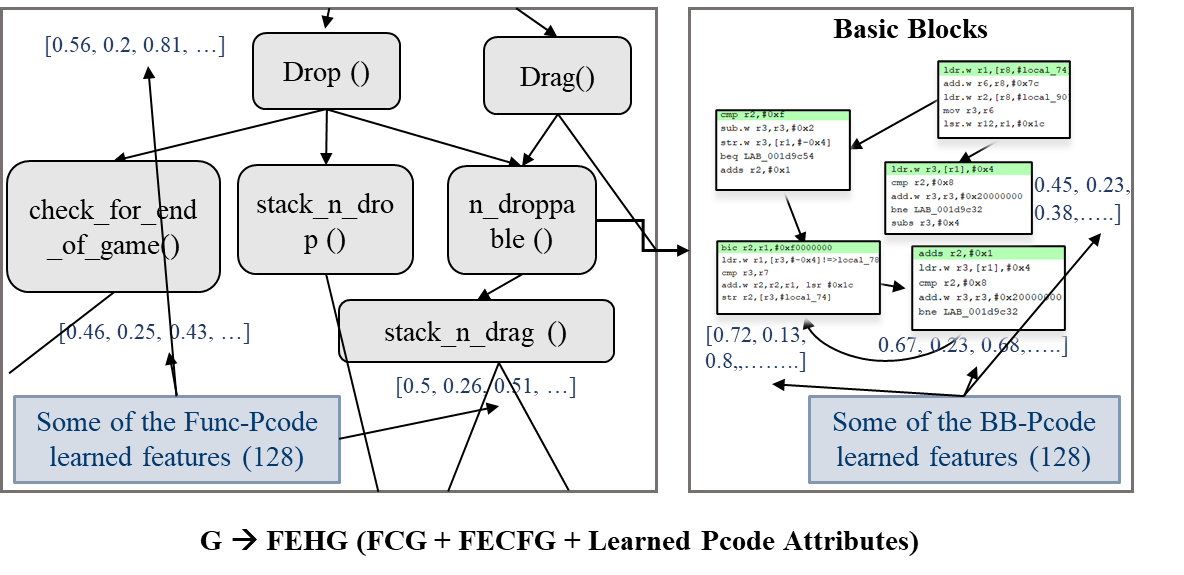}
    \caption{An example of a \textit{FEHG} of a binary.}
    \label{fig:binaryModeling}
\end{figure}

AGNOMIN introduces FEHGs, a comprehensive representation that combines CFGs, FCGs, and learned \texttt{PCode} features to enable architecture-agnostic analysis of binary functions (see Fig.~\ref{fig:binaryModeling}).
An FEHG is denoted as \(G = (V, E, \Psi)\), where $V = \{f_1, f_2, \ldots, f_n\}$  is a set of function nodes, $E$ is an adjacency matrix representing the relationships between functions, and $\Psi$ is a mapping function that assigns each function $f_i \in V$ its corresponding learned feature set $\mu_i$ extracted from \texttt{PCode}; $\Psi(f_i) = \mu_i$.

Each function node  \( f_i \) is further conceptualized as a Feature-Enriched Control Flow Graph (FECFG), represented as \( f_i = (BB, E, \psi) \). \( BB=\{bb_1, bb_2, \ldots, bb_n\} \) denotes the set of attributed basic blocks, representing the functional internal structure. \( E \) depicts the execution flow among these basic blocks, and \(\psi\) assigns each basic block \(bb_i\) an extensive set of features encapsulated in $p$-dimensional feature vector $\nu_i$ derived from \texttt{PCode}; \(\psi(bb_i) = \nu_i\). 

AGNOMIN's key technical merit lies in utilizing the multi-level perspective provided by FEHGs, capturing the internal structure of individual functions through FECFGs and their interactions with other functions through FEHG. 
This holistic view sets AGNOMIN apart from approaches that examine only CFGs or FCGs in isolation.
Moreover, AGNOMIN addresses the challenges posed by diverse architectures by employing learned features from \texttt{PCode}, an architecture-independent intermediate representation. 
By extracting features that resonate across various architectures, AGNOMIN enables accurate function name prediction even in the presence of limited debug symbols. 

\subsection{Feature Extraction from \texttt{PCode}} 
AGNOMIN enhances the binary representation by extracting rich contextual information from \texttt{PCode} sequences using a pre-trained Text-to-Text Transfer Transformer (T5) ~\cite{DBLP:journals/corr/abs-1910-10683} model.  
T5, which is renowned for its natural language processing capabilities, treats \texttt{PCode} sequences as a language, uncovering key patterns and relationships that enhance the model's understanding of binary functionalities.

\subsubsection{T5 Encoding Process} For each basic block \( bb_i \) and function \( f_i \), the associated \texttt{PCode} is input into the T5 model, transforming it into a feature vectors $\nu_i$ and  $\mu_i$ respectively:      $  \nu_i = \text{T5}(\text{\texttt{PCode}}(bb_i)) ; ~~\mu_i = \text{T5}(\text{\texttt{PCode}}(f_i)) $.
      Here, \( \text{\texttt{PCode}}(bb_i) \)  and \( \text{\texttt{PCode}}(f_i) \) represents \texttt{PCode} sequences for the basic block \( bb_i\) and \(f_i\) respectively.
\subsubsection{Why T5?}
The T5 model contains both an encoder, which consumes long documents or source language text -- and a decoder, which outputs the abstract or target-language text. The existence of both encoder and decoder significantly extends the capability of language models to handle lengthy \texttt{PCode} text strings. Decoder-only models such as GPT3~\cite{radford2018improving} lack the ability to perform feature extraction and the pre-training/self-training process. 
Encoder-only models, such as BERT~\cite{DBLP:journals/corr/abs-1810-04805}, focus on single tokens, which risks failure to capture the wider context. The pre-training of the T5 model requires this full pipeline to recover a longer sentence from corruption, and after the pre-training, the encoder part can be utilized to extract features from \texttt{PCode} strings.

\subsubsection{Why \texttt{PCode}?}
Raw binary data inherently lacks semantic meaning, making direct analysis challenging. \texttt{PCode} bridges this gap by translating assembly instructions into a more human-readable form. This allows T5, trained on text strings, to comprehend the functionalities of instructions and extract valuable features.
While we use Ghidra's \texttt{PCode}, AGNOMIN's approach can be adapted to other intermediate representations like those in IDA Pro~\cite{hex2017ida} or radare2~\cite{radare2book}. Porting AGNOMIN would primarily involve replacing the \texttt{PCode} feature extraction module with one tailored to the new IR, and potentially retraining the T5 model. The core architecture would remain largely unchanged, demonstrating AGNOMIN's flexibility and adaptability to various reverse engineering tools and workflows.

\subsubsection{Integration with FEHG} The enriched feature vectors $\nu_i$ \& $\mu_i$ are integrated into the FEHG model, updating the mapping function:
 \begin{equation}
\begin{aligned}
    \Psi(bb_i) = \text{T5}(\text{\texttt{PCode}}(bb_i)) = \nu_i ,
    \\
     \Psi(f_i) = \text{T5}(\text{\texttt{PCode}}(f_i)) = \mu_i
\end{aligned}
 \end{equation}

      for each node \( bb_i \) and
      for each node \( f_i \) in the FEHG.
This significantly improves the representation of each function and basic block, providing a deeper semantic understanding crucial for the function embedding.

\subsection{Function Embedding}

AGNOMIN employs a hierarchical graph neural network (HGNN) to process the FEHG and generate function embeddings. The HGNN consists of two key layers:

\textbf{FECFG Embedding Layer:} This layer focuses on vectorizing the control flow within each function using a Graph Convolutional Network (GCN). It computes the embedding $\phi(bb_i)$ for each basic block $bb_i$ by applying a series of graph convolutional operations. 
\toremove{FECFG considers both the control flow's structural elements and the operational semantics encoded within each basic block, synthesizing a detailed representation of the function's internal structure.}
\begin{multline}
\phi^{(l)}(bb_i) = \sigma(W^{(l)}_{\text{FECFG}} \phi^{(l-1)}(bb_i) + \\
\sum_{j \in N(bb_i)} (M^{l}\phi^{(l-1)}(bb_j)) + b^{(l)}_{\text{FECFG}})
\end{multline}
 where $\phi^{(l)}(bb_i) \in \mathbb{R}^{d_l}$ represents the embedding of basic block $bb_i$ at the $l$-th convolutional layer, $W^{(l)}_{\text{FECFG}} \in \mathbb{R}^{d_l \times d_{l-1}}$ and $b^{(l)}_{\text{FECFG}} \in \mathbb{R}^{d_l}$ are trainable weight matrix and bias vector, respectively, $\sigma$ is a non-linear activation function (e.g., ReLU), $\mathcal{N}(bb_i)$ denotes the set of neighboring blocks of $bb_i$ in the FECFG, and $M^l \in \mathbb{R}^{d_l \times d_{l-1}}$ are learnable weights.
The initial embedding $\phi^{(0)}(bb_i)$ is set to the corresponding feature vector $\nu_i$. 
After applying $L$ convolutional layers, the final embedding for each basic block $bb_i$ is obtained as $\phi(bb_i) = \phi^{(L)}(bb_i)$.

\textbf{FEHG Embedding Layer:} Implemented using a Graph Attention Network (GAT), this layer refines the function embeddings by integrating the vectorized representations from the FECFG layer with the feature vector $\mu_i$ of each function node $f_i$  in the FEHG.
The FEHG Embedding Layer computes the function embedding $\xi(f_i)$ as:
\begin{equation}
\xi(f_i) = \tanh\left(\text{GAT}\left(\theta(f_i), E_\text{CG}\right)\right)
\end{equation}
where $\theta(f_i) = \sum_{bb_j \in f_i} \phi(bb_j)$ is the aggregated embedding of all basic blocks within function $f_i$, and $E_\text{CG}$ represents the edges in the call graph (CG) of the binary. The GAT operation is defined as:
\begin{equation}
\text{GAT}(\theta(f_i), E_\text{CG}) = \sum_{f_j \in N(f_i)} \alpha_{ij} \cdot \left(W_\text{GAT} \cdot \theta(f_j)\right)
\end{equation}
where $W_\text{GAT} \in \mathbb{R}^{d_f \times d_f}$ is a trainable weight matrix, and $\alpha_{ij}$ is the attention coefficient that determines the importance of the neighboring function $f_j$ to the current function $f_i$. $\alpha_{ij}$ is computed using a self-attention mechanism:
\begin{equation}
\begin{aligned}
\alpha_{ij} = \text{softmax}_j(e{ij}) = \frac{\exp(e_{ij})}{\sum_{k \in N(f_i)} \exp(e_{ik})}
\\
e_{ij} = \text{LeakyReLU}\left(a^T \cdot \left(W_\text{GAT} \cdot \theta(f_i) | W_\text{GAT} \cdot \theta(f_j)\right)\right)
\end{aligned}
\end{equation}
where $a \in \mathbb{R}^{2d_f}$ is a trainable attention vector, and LeakyReLU is a variant of the ReLU activation function.

The concatenated embedding $x_i = [\mu_i | \xi(f_i)]$ is passed through a fully connected layer to obtain the final function embedding $\psi(f_i)$:
$
\psi(f_i) = \sigma\left(W_{fc} \cdot x_i + b_{fc}\right)
$,
where $W_{fc} \in \mathbb{R}^{d_f \times 2d_f}$ and $b_{fc} \in \mathbb{R}^{d_f}$ are learnable parameters, and $\sigma$ is a non-linear activation function.

The HGNN architecture in AGNOMIN captures both \textit{the local structural information within functions} and \textit{the global contextual information from the call graph}, enabling the model to generate architecture-agnostic function embeddings that effectively represent the semantics of binary functions across diverse architectures.
\toremove{By incorporating the attention mechanism, AGNOMIN can dynamically prioritize the information from neighboring functions based on their relevance to the current function, leading to more contextually-aware function embeddings.The concatenation of the GAT output with the original function embedding allows the model to consider both the local structural information within the function (captured by the FECFG Embedding Layer) and the global contextual information from the call graph (captured by the GAT in the FEHG Embedding Layer).
\textbf{Addressing Previous Approaches' Limitations:} Previous models, often rooted in NLP methodologies, faced the challenges of the out-of-vocabulary problem and limited contextual interpretation. \texttt{\textbf{AGNOMIN}} overcomes these by:
\begin{itemize}
    \item Embedding Contextual Relationships: It embeds not just instructions, but their broader contextual relationships within the binary.
    \item Comprehensive View: \texttt{\textbf{AGNOMIN}} integrates data from both CFGs and FCGs, providing a holistic view of each function's role.
    \item Semantic Unification: By employing IRs like \texttt{PCode}, it unifies diverse architectural instructions into a cohesive semantic framework, abstracting away syntactic differences.
\end{itemize}}

\subsection{Cross-Architecture Learning}

AGNOMIN employs a Siamese network architecture~\cite{bromley1993signature} to enable cross-architecture learning of function embeddings. 
This design choice is particularly motivated by security applications, where analyzing similar vulnerabilities or malicious code patterns across different architectures is crucial.
The Siamese network computes the distance between two function embeddings $\psi(f_i)$ and $\psi(f_j)$ as: 
$ D(\psi(f_i), \psi(f_j)) = |\psi(f_i) - \psi(f_j)|$.

The objective is to minimize this distance for pairs of functions that are known to be similar (regardless of their originating architecture) and maximize it for dissimilar pairs. This is achieved by optimizing a contrastive loss function $L$:
\begin{multline}
L(\psi(f_i), \psi(f_j), y) = y \cdot D(\psi(f_i), \psi(f_j))^2 + (1 - y) \cdot \\ 
\max(0, m - D(\psi(f_i), \psi(f_j)))^2
\end{multline}
$y$ is a binary label indicating whether the functions are similar ($y=1$) or dissimilar ($y=0$), and $m$ is a margin parameter defining the separation threshold for dissimilar embeddings.
The Siamese network optimizes this loss to guide the HGNN in generating architecture-agnostic embeddings.

\subsection{Function Matching Capability}
\label{sec:func_matching_vuln_mgmt} 
Beyond name prediction, AGNOMIN's architecture-agnostic embeddings are pivotal for \textbf{automating key aspects of software vulnerability management}, especially function matching across diverse binaries. Given a stripped binary and a reference (e.g., one with known vulnerabilities or applied patches), AGNOMIN computes function embeddings for both, using cosine similarity to identify corresponding functions. 
This capability is crucial for security-critical automated tasks. AGNOMIN can help flag potential vulnerabilities by matching functions in stripped binaries against a database of known vulnerable function embeddings. It can also precisely identifies equivalent functions in stripped, unpatched binaries for functions patched in a reference binary, guiding automated patch application or porting. 

AGNOMIN's encoder thus facilitates these automated processes by enabling direct semantic comparison between functions in vulnerable binaries and their counterparts in reference versions. The resulting similarity scores allow automated tools to pinpoint vulnerable code or confirm patch locations with high confidence, even across differing implementations. The practical effectiveness in vulnerability remediation is shown in our case study (Sec.~\ref{sec:case_study}).

\section{Function Token Prediction}
\label{sec:decoding}

\subsection{Multi-Label Classification Approach}
AGNOMIN addresses the limitations of previous approaches by formulating function name prediction as a multi-label classification (MLC)~\cite{dahiya2021deepxml, pfastrexml2016_jain, renee2023} problem, offering several advantages over previous approaches:

\textit{1) Semantic Decomposition}: Function names are decomposed into meaningful labels representing specific semantic aspects. This decomposition allows for a fine-grained and informative representation of function names, capturing nuanced relationships between different functional components.

\textit{2) Flexibility and Generalization}: MLC enables the model to generate new combinations of labels, handling unseen function names and adapting to diverse naming conventions across different software projects and architectures.

\textit{3) Handling Rare Tokens}: MLC is particularly effective in addressing the challenge of predicting low-frequency tokens within function names~\cite{dahiya2021deepxml}. By treating each semantic component as a separate label, the model can learn to predict rare tokens more effectively than sequence-to-sequence models, which often struggle with rare vocabulary items.

\textit{4) Scalability}: MLC offers improved scalability compared to multi-class classification and seq2seq models~\cite{pfastrexml2016_jain}. It can efficiently handle large label spaces without the need for an exhaustive list of all possible function names, making it well-suited for real-world applications involving diverse and extensive codebases.

AGNOMIN takes a fundamentally different approach to MLC compared to \texttt{XFL}~\cite{patrickevans2022xfl}, which uses the \texttt{PfastreXML}~\cite{pfastrexml2016_jain} library. Instead, AGNOMIN employs a deep learning-based solution that offers several advantages:

\textit{1) Deep Learning-Based Solution}: 
\texttt{PfastreXML} relies on decision tree ensembles, which can be memory-intensive when dealing with large-scale datasets~\cite{napkinxc2020jasinska, patrickevans2022xfl, prabhu2018parabel}. Additionally, the model is sensitive to the choice of hyperparameters, such as the number of trees and the maximum depth of each tree~\cite{pfastrexml2016_jain, nyre2022task}.
AGNOMIN, on the other hand, incorporates a deep learning-based decoder. This allows for more expressive modeling of the relationships between input features and output labels, capturing complex non-linear patterns~\cite{napkinxc2020jasinska, wydmuch2018no, jernite2017simultaneous}.

\textit{2) Attention Mechanisms}: The decoder architecture incorporates attention mechanisms to dynamically focus on the most relevant aspects of the input embeddings for each label prediction~\cite{vaswani2017attention}.

\textit{3) Hierarchical Label Embeddings}: AGNOMIN employs hierarchical label embeddings to navigate the vast label space efficiently, capturing label dependencies and similarities~\cite{wehrmann2018hierarchical}. 

\textit{4) Dynamic Label Pruning}: AGNOMIN implements dynamic label pruning~\cite{babbar2017dismec, babbar2019data}  to adaptively refine the set of candidate labels during inference. This technique improves both the efficiency and precision of the label prediction process, particularly for large label spaces.

Moreover, to address the challenge of imbalanced label distributions and to improve the prediction of rare labels, AGNOMIN incorporates a propensity score method similar to that used in XFL~\cite{patrickevans2022xfl} and other extreme multi-label classification approaches~\cite{dahiya2021deepxml, renee2023}.

\subsection{Robustness to Naming Convention Variability}

AGNOMIN addresses the challenge of diverse function naming conventions across software projects by analyzing the linguistic and structural patterns that define function names in binaries, including conventions like \texttt{snake\_case}, \texttt{camelCase}, and \texttt{PascalCase}.

AGNOMIN's tokenization approach builds upon the techniques introduced by \texttt{XFL}~\cite{patrickevans2022xfl}, incorporating several enhancements to address potential shortcomings. 
It begins by splitting function names using non-alphanumeric characters (e.g., underscores, hyphens) and capitalized letters as delimiters. This allows us to handle common naming conventions like \texttt{snake\_case}, \texttt{PascalCase} and \texttt{camelCase}.

However, some programmers may opt not to use any delimiters in their function names, relying instead on human contextual understanding to mentally separate words. 
For example, a function name like \texttt{"vbicacheforeachpage"} would be challenging to tokenize using only conventional delimiters. 
To address this issue, we employ a dynamic programming algorithm that splits strings into the longest group of non-overlapping substrings, each representing a recognized word from a dictionary of approximately five thousand common general-use and programming-related words. 
In the case of \texttt{"vbicacheforeachpage",} the algorithm would separate it into distinct tokens like \texttt{"vbi", "cache", "for", "each",} and \texttt{"page", } capturing the semantic components of the function name.

AGNOMIN extends this algorithm to handle cases where an exact match is not found, calculating subsequence similarities to dictionary words using forward, backward, and Levenshtein similarities. We also maintain an expanded abbreviation dictionary that maps common programming abbreviations to their full-word counterparts, accounting for verb modifying suffixes, noun pluralization, and synonyms.

After tokenization, AGNOMIN refines the label space by pruning to include only the most common N labels (typically 512 to 4,096).
This mitigates the label scarcity problem and ensures that the decoder has sufficient examples to learn from.
This simplifies the decoder's learning process and enhances its ability to generalize and make accurate predictions across diverse naming conventions. Appx.~\ref{appendix:labelDistribution} further details label sample distribution and the effects of label space pruning.

\subsection{Decoder Design}

AGNOMIN's decoder draws inspiration from Ren\'ee~\cite{renee2023}, an end-to-end multi-label classification model. However, significant modifications and enhancements have been made to the decoder design to better suit the specific challenges of architecture-agnostic function name prediction.

\subsubsection{Decoupled Encoder-Decoder Design}
Unlike Ren\'ee, which employs an end-to-end approach, AGNOMIN decouples the encoder and decoder components. 
This decision is motivated by the goal of learning architecture-agnostic features for function name prediction. 
This allows the use of the specialized hierarchical GNN~(Sec.~\ref{sec:embedding}) as the encoder, explicitly designed for learning architecture-agnostic representations of functions.

\subsubsection{Decoder Architecture} 
Ren\'ee's decoder architecture consists of a single feed-forward layer that takes the learned embeddings as input and produces label probabilities.
AGNOMIN's decoder, however, incorporates a novel attention-based head layer that precedes the feed-forward layer.

Given the learned function embeddings $\mathbf{E} \in \mathbb{R}^{N \times d}$ from the encoder, where $N$ is the number of functions and $d$ is the embedding dimension, the self-attention mechanism computes a weighted sum of the input embeddings, where the weights are determined by the similarity between each embedding and learned attention weights.
We implemented the self-attention mechanism using the \code{MultiHeadAttention}~\cite{pytorch2019} module from \textsc{PyTorch}~\cite{paszke2019pytorch}. 
This module takes the input embeddings $\mathbf{E}$ and computes the attention weights and values internally. The output is a weighted sum of the input embeddings, where the weights are learned during training to capture the most relevant information for label prediction.

\subsubsection{Training and Optimization Strategies}

The decoupled design of AGNOMIN's architecture allows for modular training, where the binary embedding can be trained separately to learn architecture-agnostic embeddings, and the decoder can then be fine-tuned for specific tasks or datasets. 
This flexibility enables AGNOMIN to leverage pre-trained embeddings and adapt the decoder to different label spaces or naming conventions, improving the overall efficiency and adaptability of the approach.
AGNOMIN uses mini-batch training to enable faster convergence and reduce memory requirements, making it scalable to large datasets~\cite{napkinxc2020jasinska, agrawal2013multi}.

To address the class imbalance problem, where some labels are significantly more frequent than others, AGNOMIN employs class-weighted loss functions. 
These functions assign higher weights to the less frequent classes, ensuring that the model pays sufficient attention to the rare labels during training~\cite{patrickevans2022xfl}.

\section{Experiment Design}
\label{sec:experiment}

\subsection{Dataset Preparation}
Our evaluation uses the ALLSTAR (Assembled Labeled Library for Static Analysis Research) dataset~\cite{ALLSTARDataset}, containing over 30,000 Debian Jessie packages pre-built for multiple CPU architectures to support software reverse engineering research.
From ALLSTAR, we curated the Diverse Architecture Binary (\textit{DAB-9k}) dataset, selecting packages with executable ELF binaries built for  \texttt{amd64}, \texttt{armel}, and \texttt{i386} CPU architectures. We chose these architectures to represent a diverse range of computing environments: \texttt{amd64} for dominant 64-bit x86 systems, \texttt{i386} for 32-bit x86 legacy systems, and \texttt{armel} for ARM-based mobile and embedded devices~\cite{Debian-GNU-Linux}. While ALLSTAR includes other architectures like \texttt{MIPS}, \texttt{PPC}, and \texttt{s390x}, we limited our selection to these three due to storage and computational constraints. However, we want to emphasize that AGNOMIN is architecture-agnostic and can process binaries from all architectures in the ALLSTAR dataset.

Our selection criteria ensured multi-architecture availability, manageable binary sizes, and the presence of ground truth symbol information. We excluded packages with binaries for only one CPU architecture, eliminated oversized packages exceeding our computational capacity, and removed binaries lacking verifiable symbol data (checked using Ghidra decompiler and Linux file command).

This process yielded the \textit{DAB-9k} dataset, comprising 9,000 ELF binaries across the three target architectures. 
Our preliminary experiments revealed that randomly splitting the dataset led to data leakage issues. To address this, we performed a non-random variant train-test split with a 9-to-1 ratio on the DAB-9k dataset. We followed the splitting method described in~\cite{xu2017neural} to ensure that binaries belonging to the same packages stay in the same set, either the training or testing set. This variant splitting method allows us to evaluate our approach more accurately.
The testing dataset, named \textit{DAB-3arch-test}, consists of \textit{DAB-amd-test}, \textit{DAB-i386-test}, and \textit{DAB-armel-test} subsets, each holding test data for its respective architecture.

Each binary in the \textit{DAB-9k} dataset undergoes an analysis using Ghidra.
Through this analysis, we extract essential information, including Ghidra's \texttt{PCode} and structural details, closely adhering to the methodology outlined in~\cite{yu2023cfg2vec}.
Specifically, we extract eight key files from Ghidra that encompass control flow data, functional call/callee relations, and \texttt{PCode} representations of the binaries. Fig.~\ref{fig:AGoG} provides a detailed illustration of this extraction process.
Table~\ref{tab:stat} in Appx.~\ref{appendix:dataset} summarizes the data statistics of the \textit{DAB} dataset, including the number of packages and binaries for each architecture, as well as the average number of function nodes, edges, and basic block (BB) nodes per binary.

\subsection{Experimental Setup}
\label{subsec:exp_setup}

\textbf{Encoder Training:} The encoder is trained separately from the decoder to avoid confounding hyperparameter effects. The objective is to find the optimal \texttt{PCode} IR setup to enhance the encoder's performance. We explore various combinations of feature extraction levels (basic block (BB), function, or both) \& number of features extracted (32, 64, 128) from each level.
The encoder is trained for 100 epochs with fixed hyperparameters to isolate the effect of the \texttt{PCode} setup.
The performance is evaluated using the precision metric, which measures the similarity between the generated embedding and the predicted similar embeddings. For each function embedding, we calculate pairwise cosine similarities with all other function embeddings and take the top list. 
If the ground truth function name is similar to the name of the predicted function embedding, it is considered a hit; otherwise, it is a miss.

\textbf{Decoder Training:}
After training the encoder, we aim to find the configuration that most effectively predicts a set of labels given a function embedding, capable of finding relevant labels for a function.
We train decoders on the embeddings generated by the best-performing model of the \texttt{PCode} configuration with various label spaces.
We evaluate the decoder's performance using various multi-label classification metrics, as described in the following subsection.
Refer to Appx.~\ref{appendix:training_setup} for details on the hardware configurations and timing benchmarks for training and inference. 

\mycomment{
To this end, we considered several metrics to evaluate the performance. F1 score is the harmonic mean of precision (ability to predict labels correctly while not predicting false labels) and recall (ability to predict labels correctly while not failing to predict true labels), making it an ideal measure for our criteria. We also consider Discounted (dCG) and Normalized Discounted Cumulative Gain (ndCG), which measure  the total relevance of the predicted labels.}

\begin{figure*}[!ht]
    \centering
    \includegraphics[trim=2pt 9pt 2pt 1pt,clip, width=0.999\linewidth]{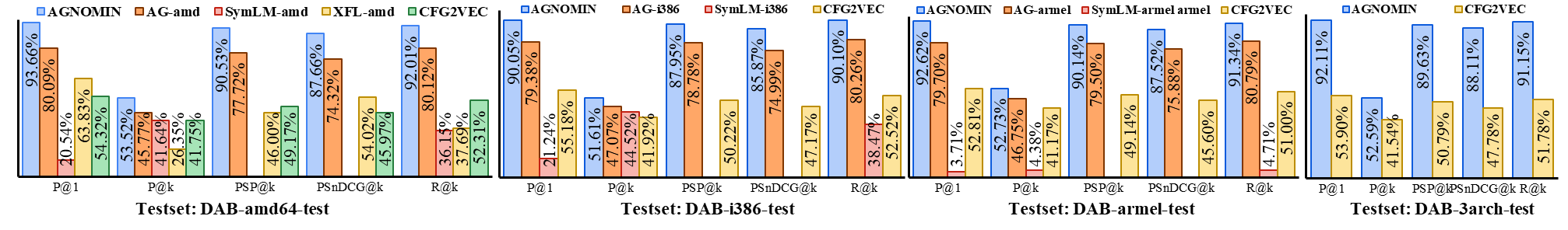}
    \caption{{Evaluation of \textit{AGNOMIN} for function name prediction against SOTA approaches across different architectures.}}
    \label{fig:RQ2andRQ3}
\end{figure*}

\subsection{Multi-Label Metrics}
We evaluate AGNOMIN using metrics
that assess its ability to identify relevant labels for each function.

\textbf{Precision @ k ($P@k$):}
identifies how many of the top-k most probable labels are contained in the ground truth label set. $P@k = \frac{1}{k} \sum_{l\in rank_k(\hat{y})}{y_l}$; 
where $rank_k(\hat{y})$ returns the k most probable labels of $\hat{y}$ in descending order \cite{Bhatia16}.

\textbf{Recall @ k ($R@k$):} identifies how many of the labels in the ground truth are contained in the $k$ most probable labels.

\textbf{Propensity-Scored Precision @ k (\textit{$PSP@k$}):} is an extension of \textit{P@k} that incorporates propensity scores to address the challenge of classifiers focusing on predicting only the most common labels~\cite{Bhatia16}. 
$
PSP@k = \frac{1}{k} \sum_{l\in rank_k(\hat{y})}{\frac{y_l}{p_l}}
$
; where $p_l$ is the propensity score of label $l$, which is proportional to the label's frequency in the training dataset.

\textbf{Propensity-Scored DCG ($PSDCG@k$) and Propensity-Scored Normalized Discounted Cumulative Gain ($PSnDCG@k$):} evaluate the decoder's ability to identify informative labels, even when these labels are relatively uncommon in the training data. PSnDCG accounts for the order of predictions, emphasizing the importance of top-ranked labels. It is derived from Discounted Cumulative Gain (DCG), which quantifies the total helpful information contained within the top $k$ predictions.
\begin{equation}
PSDCG@k = \sum_{l\in rank_k(\hat{y})}{\frac{y_l}{p_l log_2(l+1)}}
\end{equation}
\begin{equation}
PSnDCG@k = \frac{PSDCG@k}{\sum_{l\in rank_k(\hat{y})}^{k}{\frac{1}{log_2(l+1)}}}
\end{equation}

These metrics are calculated using the PyXCLib~\cite{pyxclib} library, a common tool for multi-label classification tasks.

\section{Evaluation}
\label{sec:evaluation}
To evaluate the effectiveness of AGNOMIN's approach, we conduct a comprehensive set of experiments designed to assess both its technical capabilities and practical utility. Our evaluation focuses on the following five research questions that examine different aspects of AGNOMIN's performance.

\textbf{RQ1}: What is the optimal process for embedding within AGNOMIN, and how does it perform compared to the leading architecture-independent embedding methods?

\textbf{RQ2}: Can AGNOMIN create more accurate and meaningful function names than the SOTA approaches?

\textbf{RQ3}: How does AGNOMIN's cross-architecture performance compare to other SOTA methods, indicating its suitability for diverse real-world binaries?

\textbf{RQ4}: How well can AGNOMIN generalize the learned knowledge to binaries compiled for architectures not seen during training?

\textbf{RQ5}: In what ways can the individual components of AGNOMIN be optimized to enhance its overall effectiveness?

\begin{table}[!h]
\caption{\texttt{PCode} configurations for \textit{AGNOMIN}'s encoder.}
\vspace{-1.0em}
    \centering
    \begin{tabular}{l|lllll}
    \hline
        \# Features & BB & Func & BB\_Func & CFG2VEC & ACFG  \\ \hline
        -- & - & - & - & 0.846 & 0.814\\ 
        32 & 0.885 & 0.914 & 0.933 & - & - \\ 
        64 & 0.860 & 0.933 & 0.910 & - & -  \\ 
        128 & 0.921 & 0.936 & 0.940 & - & -\\ \hline
    \end{tabular}
    \label{tab:encoderResult}
\end{table}

\subsection{RQ1: Optimal Binary Embedding}
To identify the optimal embedding process for AGNOMIN, we evaluate the encoder's performance across different \texttt{PCode} feature configurations using the precision metric as discussed in Sec.~\ref{subsec:exp_setup}. Table ~\ref{tab:encoderResult} shows the average results from training each configuration with five seeds.

Extracting 128 features per BB and/or function consistently yields higher precision than 32 or 64, suggesting richer semantic capture of basic blocks and function calls. Jointly using BB and function features (128-feature configuration) produces the most accurate embeddings by creating comprehensive contextual representations. Such high-quality embeddings are vital for precisely \textbf{identifying function correspondences, a prerequisite for tasks like patch analysis or known vulnerability detection}.

We benchmarked AGNOMIN's embeddings against \texttt{CFG2VEC}~\cite{yu2023cfg2vec}, the only architecture-agnostic function name prediction, and attributed CFG (ACFG) based models~\cite{xu2017neural}, which model control flow/call graphs and attributed control flow graphs, respectively. As shown in Table~\ref{tab:encoderResult}, AGNOMIN outperforms both, demonstrating its superiority in learning effective binary embeddings.

\begin{figure}[!b]
    \centering
    \includegraphics[trim=2pt 10pt 2pt 1pt,clip, width=0.999\linewidth]{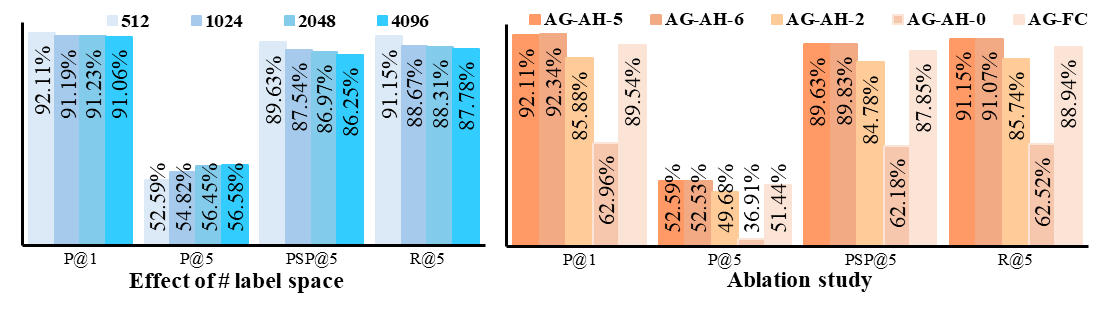}
    \caption{(left) Effect of label space size on \textit{AGNOMIN}. (right) {Comparison of \textit{AGNOMIN} and its ablated variations}.}
    
    \label{fig:labelSpaceAndAbltaion}
\end{figure}

\begin{figure*}[!t]
    \centering
    \includegraphics[trim=2pt 8pt 2pt 1pt,clip, width=0.9999\linewidth]{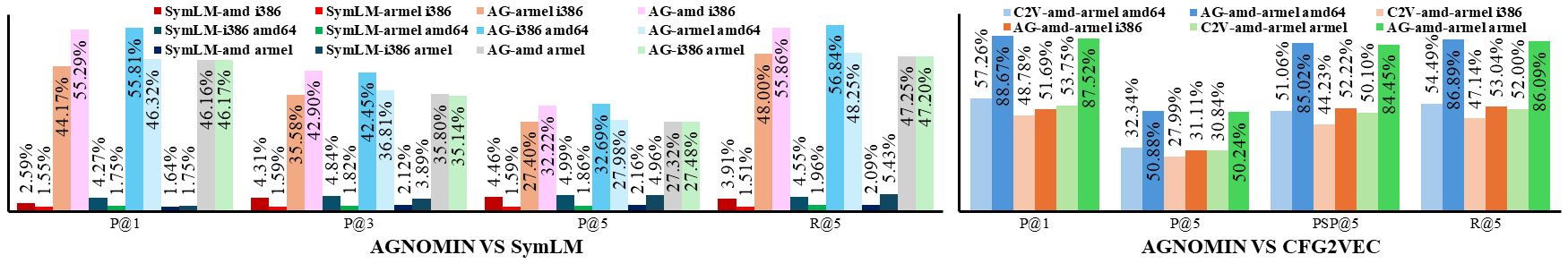}
    \caption{{Evaluation of \textit{AGNOMIN}'s ability to generalize learned knowledge compared to SOTA approaches. Each bar reflects performance for specific models, e.g., \texttt{AG-amd i386}  shows AGNOMIN's AMD-trained performance on i386 binaries.}}
    \label{fig:RQ4}
\end{figure*}

\subsection{RQ2 \& RQ3: Cross-Architecture Function Name Prediction}

To evaluate AGNOMIN's function name prediction, we generate embeddings from the encoder and randomly split them into 9:1 training/validation sets. 
We tune hyperparameters and set the decoder's optimal number of attention layers to 5 (see Sec.~\ref{subsec:ablation}).

To investigate the impact of label space size on information gain, we run experiments with four different label spaces $Y_N$ for $N = {512, 1024, 2048, 4096}$ (refer to Appx.~\ref{appendix:labelDistribution} for the label sample distribution and the impact of label space pruning). Fig.~\ref{fig:labelSpaceAndAbltaion} shows that AGNOMIN achieves higher performance on almost all reported metrics with a smaller 512-label space.
With fewer labels per function, the model can focus on learning the most informative, frequent labels, resulting in better overall performance. Moreover, increasing the label space to 4096 does not significantly reduce performance, demonstrating the benefit of hierarchical label embeddings used in AGNOMIN.
Refer to the Appx.~\ref{appendix:example_pred} for examples of AGNOMIN's prediction.

We then compare AGNOMIN with SOTA methods, including \texttt{CFG2VEC}~\cite{yu2023cfg2vec}, \texttt{SymLM}~\cite{SymLM}, and \texttt{XFL}~\cite{patrickevans2022xfl}, chosen for their performance and open-source availability. 
Other approaches are not considered due to being outperformed by the selected models~\cite{he2018debin,Gao2021,lacomis2019dire, patrickpuntrip, david2020neural}, lacking sufficient documentation for reproducibility, or requiring licensed software for dataset preprocessing~\cite{patrickpuntrip, david2020neural, asmdepictor}. 
AGNOMIN is trained in a cross-architecture setting using the \textit{DAB-9k} dataset, while separate models (\texttt{AG-amd}, \texttt{AG-i386}, and \texttt{AG-armel}) are trained for each architecture using the respective \textit{DAB-3k} datasets.

{AGNOMIN and \texttt{CFG2VEC} are architecture-agnostic, handling binaries across architectures without per-architecture training. In contrast, \texttt{SymLM} requires separate training for each target architecture, while \texttt{XFL} only supports the \texttt{x86-64}, limiting their real-world applicability across diverse architectures.}

\textbf{Comparison with \texttt{SymLM}:} \texttt{SymLM}~\cite{SymLM} combines function semantics and calling context to generate tokens representing words in a function name, using an interprocedural control flow graph (ICFG) to represent a function's context. We adapt \texttt{SymLM}'s pipeline to process our dataset and generate the required inputs.

Fig.~\ref{fig:RQ2andRQ3} presents the results of evaluating \texttt{SymLM} on our dataset, with separate models trained for each architecture (\texttt{SymLM-amd}, \texttt{SymLM-i386}, and \texttt{SymLM-armel}). 
AGNOMIN demonstrates superior performance across all architectures.
For example, on the \texttt{amd64} architecture, \texttt{SymLM-amd} achieved 41.64\% precision when predicting 5 labels, while AGNOMIN, trained on three architectures, attained 53.52\% precision (\textit{P@k, for k=5}) -- an improvement of 11.88\%. AGNOMIN also outperformed \texttt{SymLM-amd} in recall (\textit{R@k}) by 55.86\%. 
Similarly, the \texttt{AG-amd} model outperforms \texttt{SymLM-amd} by a higher margin in all metrics. On the \texttt{i386} architecture, AGNOMIN outperforms \texttt{SymLM-i386} by 51.63\% in recall. As \texttt{SymLM} does not support propensity-based scoring, we do not report those metrics for \texttt{SymLM}.
\maybe{AGNOMIN's architecture-agnostic approach, leveraging graph-based representations and learned \texttt{PCode} features, enables it to generate precise function name predictions, outperforming \texttt{SymLM}'s ICFG-based approach.}

\textbf{Comparison with \texttt{XFL}:} \texttt{XFL}~\cite{patrickevans2022xfl}
uses static analysis-based features with local context from call graphs and global context from the binary to create a combined embedding of a function's semantics. 
It employs PfastreXML to predict the probability of each label in the label space. We trained \texttt{XFL} on our \texttt{amd64} dataset using 512 labels, as \cite{patrickevans2022xfl} suggests better results with smaller label spaces.

As shown in Fig.~\ref{fig:RQ2andRQ3}, AGNOMIN outperforms \texttt{XFL} across all metrics on \texttt{amd64} binaries, achieving \textit{29.83\%} higher \textit{P@1, 27.17\%} higher \textit{P@k (k=5), 54.32\%} higher \textit{recall}, \textit{44.53\%} higher \textit{PSP}, and \textit{33.64\%} higher \textit{PSnDCG}. \texttt{AG-amd} also surpasses \texttt{XFL} by a higher margin in all metrics. 
These demonstrate AGNOMIN's ability to identify informative labels for functions, even when these labels are relatively uncommon in the training data. AGNOMIN's higher propensity-based scores compared to \texttt{XFL} show that our approach effectively addresses the tail-label prediction problem. 
\texttt{XFL}'s reliance on \texttt{PfastreXML} and decision tree ensembles hinders its ability to capture complex non-linear relationships between input features and output labels.

\textbf{Comparison with \texttt{CFG2VEC}:} \texttt{CFG2VEC} uses control-flow and function-call graphs from Ghidra to create a comprehensive binary representation using a dual-layered Graph-of-Graph representation. We compare AGNOMIN with \texttt{CFG2VEC} by setting up the necessary environment according to their GitHub repository instructions.

Since \texttt{CFG2VEC} predicts full function names by comparing its embeddings with other functions in its database, a direct comparison with AGNOMIN's multi-label prediction approach is not feasible. To address this, we devised two methods for comparing the approaches:
(1) We used AGNOMIN's encoder to generate function embeddings and incorporated them into \texttt{CFG2VEC}'s pipeline for function name prediction. This allowed us to compare the effectiveness of the two approaches in predicting function names based on their respective embeddings. 
As discussed in Sec.~\ref{subsec:exp_setup}, we use the precision metric (a multi-class classification metric) where for each function embedding, if the ground truth function name is similar to the name of the predicted function embedding, it is considered a hit; otherwise, it is a miss.
The results are presented in Table~\ref{tab:cfg2vec-agnomin}. Across all architectures, AGNOMIN outperformed \texttt{CFG2VEC} by 24.48\% in predicting full function names. 
(2) We incorporated \texttt{CFG2VEC} as the function encoder in AGNOMIN's architecture, replacing AGNOMIN's encoder while using Ren\'ee as a decoder. This enabled us to evaluate the impact of \texttt{CFG2VEC}'s embeddings on multi-label prediction performance (see Fig.~\ref{fig:RQ2andRQ3}).

AGNOMIN outperformed \texttt{CFG2VEC} in both evaluation methods. For example, using \texttt{CFG2VEC} as an encoder results in {21.47\%} lower precision (\textit{P@5}) and 39.37\% lower recall across all architectures (\texttt{amd64}, \texttt{armel}, and \texttt{i386}). 
AGNOMIN's FEHGs provide a more comprehensive representation by incorporating learned \texttt{PCode} features, allowing it to capture more relevant semantic information and generate more accurate function name predictions.

\begin{table}[!h]
    \centering
    \caption{Performance of \texttt{CFG2VEC} and \textit{AGNOMIN'}s encoder.}
    \begin{tabular}{ c| c c c c }
        \hline
        ModelName & \texttt{amd64} & \texttt{armel} & \texttt{i386} & 3arch  \\
        \hline
       \texttt{CFG2VEC}      & 69.11\% & 69.41\% & 70.66\% & 69.75\% \\
       \textit{AGNOMIN\_Encoder} & 91.55 \% & 92.19 \% & 91.76\% & 94.23\%  \\
        \hline
    \end{tabular}
    \label{tab:cfg2vec-agnomin}
\end{table}
\subsection{RQ4: Generalizing the Learned Knowledge}

Fig.~\ref{fig:RQ4} shows how well AGNOMIN transfers knowledge across architectures.
Existing models like \texttt{SymLM} struggle significantly when analyzing binaries from architectures they weren't trained on. For example, when we take the \texttt{SymLM} model trained only on \texttt{amd64} architecture (\texttt{SymLM-amd}) and test it on i386 binaries, its recall rate drops dramatically to just 3.91\%. In contrast, our AGNOMIN model trained on the same \texttt{amd64} architecture (\texttt{AG-amd}) only achieves a much higher 47.25\% recall rate when tested on unseen \texttt{i386} binaries. This highlights AGNOMIN’s strong cross-architecture generalization, even when it has never seen the target architecture during training.

To further evaluate this generalization capability, we conducted a more comprehensive experiment. We trained both \texttt{CFG2VEC} and AGNOMIN using only binaries from \texttt{amd64} and \texttt{armel} architectures (creating models we call \texttt{C2V-amd-armel} and \texttt{AG-amd-armel}), and then tested them on binaries from the \texttt{i386} architecture --- which neither model had seen during training.
AGNOMIN showed robust performance under this challenge: despite never seeing \texttt{i386} binaries during training, \texttt{AG-amd-armel} achieved 50\% precision and 53.04\% recall. To put these numbers in perspective, when compared to \texttt{C2V-amd-armel}'s performance, AGNOMIN showed consistently better results across all metrics: it performed better by 2.92\% in P@1, 3.1\% in P@5, and 5.89\% in recall.
When we look at the bigger picture across all three architectures (\texttt{AMD64}, \texttt{ARMEL}, and \texttt{i386}), the advantages of AGNOMIN become even more apparent. \texttt{AG-amd-armel} outperforms \texttt{C2V-amd-armel} by substantial margins: 25.56\% higher in precision (P@5) and 35.51\% higher in recall. These numbers indicate that AGNOMIN not only handles unseen architectures better but also maintains stronger overall performance across both familiar and new architectures.
The practical advantage of this generalization capability is demonstrated in Sec.~\ref{subsec:case_study_nasa}, where AGNOMIN successfully analyzed an AARCH64 v8A binary without any prior training on this architecture.

\subsection{RQ5: Ablation Study}
\label{subsec:ablation}
We conducted an ablation study to examine each component's contribution to the overall effectiveness of AGNOMIN.

First, we examined the impact of \texttt{PCode} feature configurations on the model's performance. Table~\ref{tab:bbFuncFeatures} shows how different feature extraction levels affect prediction accuracy. The results demonstrate that extracting 128 features per basic block and/or function consistently yields higher precision than 32 or 64 features, with the 128-feature configuration achieving 92.11\% P@1 and 91.15\% R@5. This suggests that more features better capture the semantics of basic blocks and function calls, enabling more accurate predictions.

\begin{table}[!ht]
    \centering
    \caption{Effects of features extracted from \texttt{PCode}.}
    \begin{tabular}{ c| c c c c c}
        \hline
        \# BB\_Func  & P@1 & P@5 & PSP@5 & nDCG@5 & R@5  \\
        \hline
       128 & 92.11\% & 52.59\% & 89.63\% & 90.91\% & 91.15\% \\
       64  & 91.96\% & 53.19\% & 91.20\% & 91.13\% & 91.22\% \\
       32  & 90.12\% & 52.58\% & 89.06\% & 88.86\% & 89.02\% \\
        \hline
    \end{tabular}
    \label{tab:bbFuncFeatures}
\end{table}

Next, we varied the number of attention heads in the decoder layer, training \texttt{AG-AH-N} models with N attention heads (N = 0, 2, 4, 6). Fig.~\ref{fig:labelSpaceAndAbltaion} shows that as the number of attention heads increases, the model's performance improves. However, we chose 5 as the default number of attention heads, as the performance gain from 5 to 6 was not significant.

We also modified the encoder model by changing the GAT layer to a fully connected layer, creating \texttt{AG-FC}. The results indicate that modifying the GAT layers significantly affects performance. 
{These highlight the importance of the decoder's attention mechanism and the encoder's GAT layer. The attention mechanism enables the decoder to focus on the most relevant aspects of the input embeddings for each label prediction.
Similarly, the GAT layer is crucial in refining the function embeddings by integrating the vectorized representations from the FECFG layer with the feature vectors of each function node in the FEHG.}

\section{Case Studies: Enabling Automated Analysis and Remediation}
\label{sec:case_study}
While quantitative evaluations demonstrate AGNOMIN's technical proficiency, its tangible value in addressing real-world security challenges, particularly in \textit{automating aspects of vulnerability analysis and patching}, is best illustrated through practical applications. This section presents two case studies from a DARPA-sponsored hackathon, showcasing how AGNOMIN provides essential semantic context to aid reverse engineers and \textit{enable more effective automated approaches} in scenarios involving stripped, multi-architecture binaries. AGNOMIN is integrated into the Ghidra reverse engineering framework as a plugin, facilitating the reconstruction of symbol names and functional context crucial for these tasks.

\subsection{Case Studies from DARPA Hackathon Context}
\label{subsec:darpa_hackathon_context}
Evaluated in a DARPA hackathon~\footnote{This subsection is heavily anonymized for review.}, AGNOMIN tackled challenges requiring finding and patching vulnerabilities in C/C++ binaries across x86, ARM, and PowerPC. In these time-critical scenarios, AGNOMIN bridged the semantic gap from stripped symbols, \textit{accelerating the vulnerability discovery and patching pipeline}.

\subsubsection{Pinpointing Functions for Targeted Binary Modification}
\label{subsec:case_study_nasa}
One challenge involved a simulated NASA Mars rover whose control binary required a micro-patch to alter its behavior and call a specific ``sharpen" function. While a functional patch, the process of precisely identifying the target function within the stripped binary mirrors the critical first step in \textit{automated security patching}: locating the exact code segment requiring modification.

Reverse engineers (REs) utilized AGNOMIN to rapidly gain semantic understanding of the stripped binary. By leveraging a reference binary, they employed AGNOMIN's function matching to identify likely image processing functions. This, combined with function name prediction for broader context, enabled them to accurately pinpoint the function needing the patch. This targeted identification, significantly accelerated by AGNOMIN, is crucial for \textit{reducing the manual effort typically required before automated or semi-automated patching tools can be effectively applied}.

\begin{figure}[!ht]
    \centering
    \includegraphics[width=.9\linewidth]{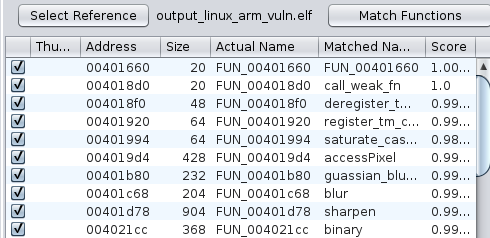}
    \caption{AGNOMIN's Function Matching performance on stripped NASA Rover \texttt{AARCH64} binary, demonstrating its utility in identifying patch targets.}
    \label{fig:matchingTable}
\end{figure}

\textbf{Function Matching for Patch Targeting:} AGNOMIN's encoder matched functions in the stripped \texttt{AARCH64 v8A} C++ rover binary against a reference with symbols. It correctly identified 51 of 64 (79.7\%) target functions via direct matching (Fig.~\ref{fig:matchingTable}), increasing to 92.1\% with the top 3 matches. This high accuracy on an \textit{unseen architecture and codebase without retraining} underscores AGNOMIN's robustness, a key requirement for \textit{scalable automated systems needing to analyze and facilitate patching across diverse binaries}.

\textbf{Function Name Prediction for Semantic Clues:} Without a reference, name prediction offered semantic clues, guiding REs (and potentially \textit{automated heuristics}) to relevant functions, narrowing the search for patch points and aiding \textit{automated decision-making}.

\subsubsection{Accelerating Security Vulnerability Remediation}
\label{subsec:case_study_crypto}
A challenge involved updating deprecated cryptography in a complex, optimized, and statically linked binary, where \textit{automated identification of vulnerable code is difficult}. AGNOMIN's function matching was key, efficiently pinpointing matches between the stripped, vulnerable binary and a secure reference. This drastically \textit{reduced the search space for REs and potential automated tools}, especially where traditional analysis fails. Its architecture-agnosticism ensured reliable matching, enabling \textit{consistent, automated security analysis and patch preparation}.

\subsection{Impact on Reverse Engineering Practices}
\label{subsec:impact_binary_analysis}
These case studies show AGNOMIN's utility in contributing to \textit{automated vulnerability detection and patching}.
By recovering semantic context in stripped binaries, it reduces analysis time and, crucially, provides a foundation for \textit{enhancing automated security tools}. Its cross-platform capabilities extend the usability of data-driven tools to rare or new architectures without extensive retraining. The goal is to collect an artifact/embedding database for a large corpus of source code, compilers, and architectures, further enhancing the usefulness of tools like AGNOMIN.

\section{Conclusion}
\label{sec:discussion}

\subsection{ Challenges and Limitations}

While AGNOMIN makes significant advancements in architecture-agnostic function name prediction, several challenges and limitations need addressing. 
Binary obfuscation techniques designed to hinder reverse engineering pose hurdles, necessitating more robust handling methods. 
Moreover, domain-specific naming conventions across programming languages, frameworks, and application domains may not be well-represented in the training data, requiring broader domain-specific datasets and more flexible naming convention adaptation techniques. 

Compiler optimizations pose another challenge. Varying optimization levels can change the structure and characteristics of binary code, which may impact AGNOMIN's performance. Although AGNOMIN's architecture-agnostic approach lays the groundwork for addressing these variations, we did not explicitly assess its performance in this study. We made this decision to stay focused on the primary challenge of cross-architecture generalization and due to the computational resources needed for such a comprehensive evaluation. However, AGNOMIN's adaptable design positions it well to address this challenge in a manner similar to how it handles different architectures.

\subsection{Conclusion}
AGNOMIN presents an effective solution for architecture-agnostic function name prediction in stripped binaries using FEHGs, attention-based decoding, and algorithmic enhancements. 
Evaluations on 9,000 ELF binaries across three architectures demonstrate AGNOMIN's superior performance, with improvements of up to {27.17\%} in precision compared to~\cite{patrickevans2022xfl} and {11.88\%} in precision and 55.86\% in recall compared to~\cite{SymLM} in precision. 
AGNOMIN also exhibits strong generalization capabilities on unseen architectures.
The open-source release of AGNOMIN as a Ghidra plugin facilitates its adoption by the reverse engineering community, and case studies showcase its potential to streamline reverse engineering workflows.

\section{Acknowledgement}
This material is based upon work supported by the Defense Advanced Research Projects Agency (DARPA) and Naval Information Warfare Center Pacific (NIWC Pacific) under Contract Number N66001-20-C-4024.
The views, opinions and/or findings expressed are those of the author and should not be interpreted as representing the official views or policies of the Department of Defense or the U.S. Government.
\bibliographystyle{IEEEtran}
\bibliography{bibfile}
\appendices
\begin{table*}[!h]
    \centering
    \caption{The statistics of DAB dataset.}
\begin{tabular}{lccccccc}
\hline
{Archi} & {Packages} & {Binaries} & {Functions} & {Edges in FEHG} & {Basic Blocks} & {Edges in FECFG}  \\ \hline
\texttt{armel} & 351 & 3000& 133341   & 226252 & 1733639 & 2409644 \\
\texttt{i386}  & 351 & 3000& 135925   & 263099 & 1555106 & 2170903   \\
\texttt{amd64} & 351 & 3000& 126830   & 222203 & 1557687 & 2179100  \\
\textit{DAB}   & 1053& 9000& 396096   & 711554 & 4846432 & 6759647  \\ \hline 
\end{tabular}
\label{tab:stat}
\end{table*}

\section{More on Experimental Design}

\subsection{Dataset Statistics}
\label{appendix:dataset}
The Diverse Architecture Binary (DAB) dataset, used for training and evaluating AGNOMIN, comprises a total of 9,000 ELF binaries across three architectures: \texttt{\texttt{\texttt{amd64}}}, \texttt{armel}, and \texttt{i386}. Table~\ref{tab:stat} summarizes the data statistics of the \textit{DAB} dataset, including the number of packages and binaries for each architecture, as well as the average number of function nodes, edges, and basic block (BB) nodes per binary.
These statistics provide insights into the scale and complexity of the dataset, highlighting the diverse range of binaries and architectures covered. The substantial number of function nodes, edges, and basic block nodes underscores the importance of AGNOMIN's ability to effectively represent and analyze these complex structures, further emphasizing the significance of the Feature-Enriched Hierarchical Graph (FEHG) approach.

\subsection{Training Setup}
\label{appendix:training_setup}
We conducted all experiments on a server equipped with Intel Core i7-7820X CPUs @ 3.60GHz and 16GB RAM. To leverage the computational power of GPUs for model training and inference, we utilized three different GPU configurations: NVIDIA Tesla V100, A30 Tensor Core GPU, and A100 Tensor Core GPU.

Table~\ref{tab:time} shows the approximate time required for training and inference tasks on each GPU configuration. As can be observed, the training process for AGNOMIN's models can be computationally intensive, taking several hours depending on the GPU configuration. However, once trained, the inference time for function label prediction on a single binary is remarkably fast, taking only a minute or less.

This efficient inference time is a crucial factor in ensuring AGNOMIN's practical utility as a reverse engineering tool. When integrated into frameworks like Ghidra, AGNOMIN can provide near real-time function name predictions without introducing significant delays in the reverse engineering workflow. This seamless integration allows reverse engineers to leverage AGNOMIN's capabilities effectively, enhancing their productivity and efficiency in analyzing stripped binaries.

\begin{table}[!ht]
    \centering
    \caption{Time taken for training and inference tasks on three different GPU configurations. The inference is made on one binary.}
    \vspace{-0.5em}
    \begin{tabular}{  c| c | c | c  }
        \hline
           Model  &  Tasks  & Compute Resource & Runtime\\ \hline
\multirow{7}{*}{Encoder} & \multirow{3}{*}{Training}  & V100 &  46h 03m 16s \\ \cline{3-4} 
 &    & A30  &  35h 11m 0s \\ \cline{3-4} 
 &    & A100 &  32h 17m 58s \\ \cline{2-4} 
 & \multirow{4}{*}{Inference} & V100 & 0.575s   \\ \cline{3-4} 
 &    & A30  &  0.445s \\ \cline{3-4} 
 &    & A100 &  0.6563s \\ 
 \hline
\multirow{7}{*}{Decoder} & \multirow{3}{*}{Training}  & V100 &  13h 26m 38s \\ \cline{3-4} 
 &    & A30  &  8h 56m 49s \\ \cline{3-4} 
 &    & A100 &  8h 55m 36s \\ 
 \cline{2-4} 
 & \multirow{4}{*}{Inference} & V100 &  42.881 \\ \cline{3-4} 
 &    & A30  & 41.881  \\ \cline{3-4} 
 &    & A100 & 36.881  \\ 
 \hline
    \end{tabular}
    \vspace{-1.0em}
    \label{tab:time}
\end{table}

\subsection{Label Distribution}
\label{appendix:labelDistribution}
\begin{figure}[!ht]
    \centering
    \includegraphics[clip, width=.99\linewidth]{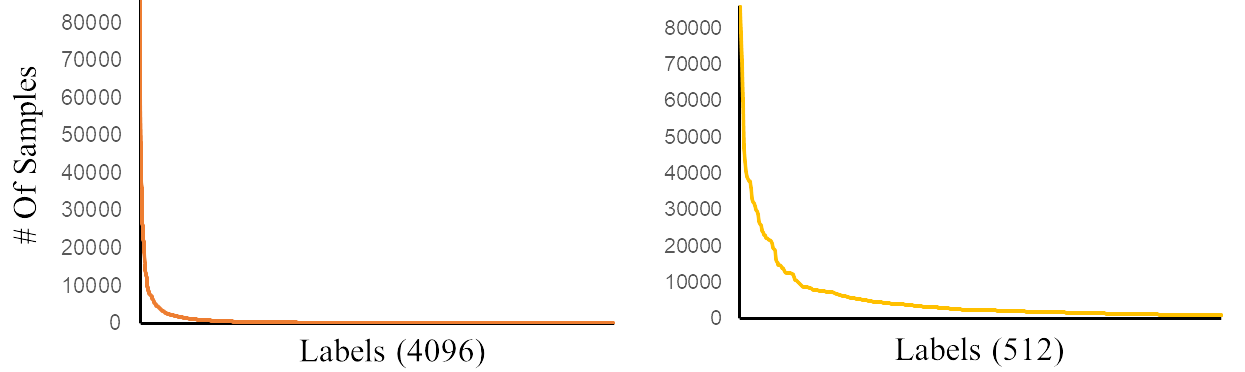}
    \caption{Label distribution.}
    \label{fig:labelProfile}

\end{figure}

Figure~\ref{fig:labelProfile} (left) shows the distribution of function labels in the \textit{DAB} dataset when considering the entire label space of 4096 labels. As evident from the figure, there is a significant class imbalance, where some labels appear frequently while others occur only a few times. This imbalance is a common challenge in multi-label classification tasks, as models can tend to focus on predicting the most common labels, potentially neglecting the rare but informative labels.

To mitigate the label scarcity problem and ensure that the decoder has sufficient examples to learn from, AGNOMIN employs a label space pruning strategy. By retaining only the N most common labels, with N typically set in the range of 512 to 4,096, AGNOMIN focuses on the most frequently occurring labels, reducing the impact of rare labels on the training process.
Figure~\ref{fig:labelProfile} (right) shows the label distribution after pruning the label space to the 512 most common labels. This pruning has lessened the imbalance in the label distribution. 

To address the remaining imbalance and the tail-label prediction problem, AGNOMIN employs propensity-based metrics, such as Propensity-Scored Precision (PSP), Propensity-Scored Discounted Cumulative Gain (PSDCG), and Propensity-Scored Normalized Discounted Cumulative Gain (PSnDCG). These metrics incorporate propensity scores that are inversely proportional to the label's frequency in the training dataset, effectively rewarding the model for successfully predicting rare labels.

\section{More Experimental Results}

\begin{table*}[!h]
    \centering
    \caption{Some examples of AGNOMIN's predictions.$^{\mathrm{*}}$}
    \vspace{-0.5em}
\begin{tabular}{lllll|lllll}
\hline
\multicolumn{5}{c|}{Predicted$^{\mathrm{1}}$}                                                                                                        & \multicolumn{5}{c}{Ground Truth$^{\mathrm{2}}$}                                                      \\ \hline
multiplication & information                 & one                         & nr                          & resolve                   & resolve                    & nr  & one      & multiplication & information \\ \hline
multiplication & pf    &       &       &     & pf   & multiplication &          &     &  \\ \hline
cs  & clone & {\color[HTML]{FE0000} so}   & {\color[HTML]{FE0000} main} &     & cs   & clone          & so       &     &  \\ \hline
register       & {\color[HTML]{FE0000} tm}   & {\color[HTML]{FE0000} get}  & test  & cs  & test & cs  & clone    &     &  \\ \hline
so  & {\color[HTML]{FE0000} main} &       &       &     & so   &     &          &     &  \\ \hline
pt  & wrap  & type  & res   &     & res  & pt  & type     & wrap&  \\ \hline
read& {\color[HTML]{FE0000} new}  & {\color[HTML]{FE0000} quit} &       &     & read &     &          &     &  \\ \hline
ae  & weak  & vi    & buffer& unknown        & ae   & weak& unknown  & vi  & buffer      \\ \hline
register       & tm    & get   & {\color[HTML]{FE0000} test} & {\color[HTML]{FE0000} cs} & {\color[HTML]{3166FF} bin} & tm  & register & get &  \\ \hline
register       & tm    & get   & {\color[HTML]{FE0000} test} & {\color[HTML]{FE0000} cs} & tm   & register       & get      &     &  \\ \hline
lib & string& in    & hmac  & fin & in   & string         & fin      & hmac& lib         \\ \hline
key & dumb  &       &       &     & dumb & key &          &     & 
\\
\hline
\multicolumn{10}{l}{$^{\mathrm{*}}$ More examples can be found in the supplementary materials.}\\
\multicolumn{10}{l}{$^{\mathrm{1}}$ Words written in {\color[HTML]{FE0000} red} are predicted by the model but not found in the ground truth.}\\
\multicolumn{10}{l}{$^{\mathrm{2}}$ Words written in {\color[HTML]{3166FF} blue} are found in the ground truth but not predicted by the model.}\\
\
\end{tabular}
\label{tab:prediction_examples}
\end{table*}

\subsection{Performance of \texttt{CFG2VEC} when used as an encoder}
\label{appendix:cfg2vec}

Table~\ref{tab:cfg2vecEncoder} shows how \texttt{CFG2VEC} performs as an encoder in a multi-label function name prediction setup using Ren\'ee's as a decoder model. 
\begin{table}[!h]
    \centering
    \caption{Performance of \texttt{CFG2VEC} when used as an encoder.}
    \begin{tabular}{ c| c c c  c}
        \hline
        Test Dataset  & P@1 & P@5 & PSP@5 &  R@5  \\
        \hline
       \textit{DAB-amd-test} & 54.32\% & 31.15\% & 49.17\% &  52.31\% \\
       \textit{DAB-i386-test } & 55.18\% & 31.86\% & 50.22\% &  52.52\% \\
       \textit{DAB-armel-test } & 52.81\% & 30.60\% & 49.14\% & 51.00\% \\
       \textit{DAB-3arch-test}  & 53.90\% & 31.12\% & 50.79\% & 51.78\% \\
        \hline
    \end{tabular}

    \label{tab:cfg2vecEncoder}
\end{table}

Table~\ref{tab:cfg2vec2archEncoder} shows how \texttt{CFG2VEC-amd-armel}, \texttt{CFG2VEC} trained with binaries compiled for \texttt{amd64} and \texttt{armel} architectures, performs as an encoder in a multi-label function name prediction setup using Ren\`ee's as a decoder model. 

\begin{table}[!h]
    \centering
    \caption{Performance of \texttt{\texttt{CFG2VEC-amd-armel}} when used as an encoder.}

    \begin{tabular}{ c| c c c  c}
        \hline
        Test Dataset  & P@1 & P@5 & PSP@5 &  R@5  \\

        \hline
       \textit{DAB-amd-test} & 57.26\% & 32.34\% & 51.06\% & 54.49\% \\
       \textit{DAB-i386-test } & 48.78\% & 27.99\% & 44.23\% & 47.14\% \\
       \textit{DAB-armel-test } & 53.75\% & 30.84\% & 50.10\% & 52.00\% \\
       \textit{DAB-3arch-test}  & 53.38\% & 30.49\% & 49.91\% & 51.38\% \\
        \hline
    \end{tabular}

    \label{tab:cfg2vec2archEncoder}
\end{table}

\subsection{Examples of AGNOMIN's Predictions}
\label{appendix:example_pred}
Some examples of AGNOMIN's function label predictions are depicted in Table~\ref{tab:prediction_examples}.

\end{document}